\documentclass[iop]{revtex4-1}

\usepackage{amsmath}
\usepackage{amssymb,amsfonts,latexsym}
\usepackage{bm}
\usepackage[mathcal]{euscript}
\usepackage{graphicx}
\usepackage{epsfig}
\usepackage{color}

\newcommand{\beq}{\begin{equation}} 
\newcommand{\eeq}{\end{equation}}  

\begin{document}

\title{A statistical physics viewpoint on the dynamics of the bouncing ball} 

\author{Jean-Yonnel Chastaing}
\affiliation{Laboratoire de Physique, Ecole Normale Sup\'erieure de Lyon,
CNRS, Universit\'e de Lyon, 46 All\'ee d\'\,Italie, 69364 Lyon Cedex, France.} 

\author{Eric Bertin}
\affiliation{Universit\'e Grenoble Alpes, LIPHY, F-38000 Grenoble, France.}
\affiliation{CNRS, LIPHY, F-38000 Grenoble, France}

\author{Jean-Christophe G\'eminard}
\affiliation{Laboratoire de Physique, Ecole Normale Sup\'erieure de Lyon,
CNRS, Universit\'e de Lyon, 46 All\'ee d\'\,Italie, 69364 Lyon Cedex, France.}

\begin{abstract}
We study from a statistical physics perspective the dynamics of a bouncing ball maintained in a chaotic regime thanks to collisions with a plate experiencing an aperiodic vibration. We analyze in details the energy exchanges between the bead and the vibrating plate, and show that the coupling between the bead and the plate can be modeled in terms of both a dissipative process and an injection mechanism by an energy reservoir, where the dynamics of the reservoir obeys only a 'blurred' version of detailed balance. An analysis of the injection statistics in terms of fluctuation relation is also provided.
\end{abstract}

\maketitle

\section{Introduction}

During the last decades, physicists have developed several tools to study the statistical properties of physical systems violating
the requirements of standard statistical mechanics of thermal equilibrium.
Such systems are ubiquitous in nature. 
For instance, the dynamics of the system can be so slow that the thermodynamic equilibrium is, in practice, never reached
as is the case for materials below the glass transition \cite{Berthier11}.
Conversely, some systems may require a continuous injection of energy to be maintained in a steady-state. This energy injection may occur in the bulk, like in active systems, or from the boundary as is typical in shaken or sheared materials.
Bulk driven systems can be found for instance in biology where, due to improvements of observation techniques, the fluctuations of tiny active units
like molecular motors can now be studied \cite{Prost09,Seifert12,Marchetti13}.
On the other side, granular media are a typical example of an externally driven system that, due to the dissipative nature of the contact between the grains, cannot sustain a fluctuating state in the absence of driving. Such granular systems cannot be described by the classical tools of thermodynamics \cite{Jaeger92}. 

Model granular media experiments have become a convenient means of investigating the physics of non-equilibrium steady state (NESS) systems \cite{Crooks99,Gallavotti95,Evans02,Seifert05,Kurchan98}.
Specially, the interest in the effects of the dissipative nature of the collision between the particles led to the experimental realization of granular gases obtained, in three-dimensions (3D), by vibrating the container \cite{Falcon99,Falcon06,Naert12} and, in two-dimensions (2D), by considering beads covering partially a horizontal surface animated by a vertical motion \cite{Olafsen98}.
All these experiments, together with their theoretical counterpart, revealed deviations from the equilibrium statistics, especially in the velocity distributions \cite{Losert99}. 

At this point, it is particularly interesting to raise the question of the thermalization of the system.
Indeed, especially in the case of 2D experiments, one can wonder if the deviation of the velocity statistics from the equilibrium ones comes from the dissipative nature of the collisions between the particles or from the peculiar nature of the thermostat the beads are in contact with. Let us for instance remark that, for a dissipative gas in contact with a thermostat, a decrease of the kinetic temperature is expected from an increase of the density (because of the increase of the collision rate) \cite{Huntley98}. In the experiments, one rather observes an increase of the temperature with the density \cite{Losert99,Geminard04}. It is thus natural to question the quality of the thermostat constituted by the vibrating plate in the 2D experiments.

Thinking of the most simple realization of a system in contact with the vibrated plate, 
one naturally comes to the problem of the bouncing bead which has been widely studied
during the late 1980's to investigate the route to chaos by period doubling \cite{Holmes82,Everson86}. 
In the chaotic regime, the system can however be regarded as a single particle in contact with an energy reservoir
and one can instead focus on the statistical properties of the energy of the bead, like its average value and its fluctuations. 
In their experimental realization of the problem, using a sinusoidal motion of the plate,
Warr {\it et al} observed that the probability distribution function (PDF) of the particle
velocity is well-approximated by a Gaussian distribution, and the PDF of the particle energy by a Boltzmann-type distribution \cite{Warr96}. However, a detailed study of the correlation of the energy states
through the collisions reveals partial synchronization of the bead motion with that of the vibrating surface, even if the bead experiences a chaotic trajectory \cite{Chastaing15}. 
In addition, in another study, it has been shown that the removal of part of the synchronization with the plate motion, can lead to an increase of the average energy of the particle for the same characteristics of the plate vibration \cite{Geminard03}, questionning again the notion of thermostat.

Here, we study from a statistical physics perspective the dynamics of a bouncing particle maintained in a chaotic regime thanks to collisions with a plate experiencing an aperiodic vibration. 
Indeed, we proposed lately a manner to produce an aperiodic vibration of the plate while keeping the vibration characteristics, especially its maximum velocity, well-defined. We thus propose an alternative way to produce a random vibration of the plate which differs significantly from the random vibration used in some recent experiments \cite{Garcia-cid15}. 
We first show that the height, velocity and energy distributions resemble the equilibrium distributions \cite{Warr96}, leading to a measure of the particle temperature. Then, we analyze in details the energy exchanges between the bead and the vibrating plate, and show that these energy exchanges
can be modeled in terms of both a dissipative process and an injection mechanism by an energy reservoir. The latter partition of the energy exchanges makes possible to recover the minimum reversibility (any transition from one value of the energy to another is possible in reverse direction) that makes possible the use of several tools of non-equilibrium thermodynamics. We thus subsequently estimate the rate of entropy production, test the detailed balance and evaluate the validity of fluctuation relations, each of these tools leading to the introduction of its own temperature scale, that can be compared one to the other.

\section{Description of the system}

\begin{figure}[h]
\begin{center}
\includegraphics[width=.35\columnwidth]{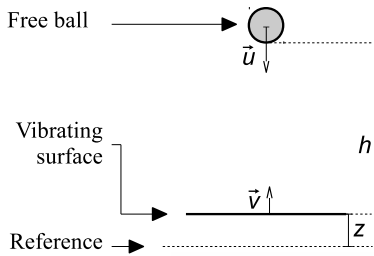}
\end{center}
\caption{\label{fig:setup} Sketch of the system.}
\end{figure}

The system under study consists of a bead bouncing freely upon a horizontal plate which oscillates along the vertical direction (Fig.~\ref{fig:setup}). 
We denote by $z(t)$ the height of the plate and by $v(t) \equiv dz/dt$ its velocity. In the same way, $h(t)$ and $u(t) \equiv dh/dt$ are the height and the velocity of the bead. We choose a common height reference, in such a way that $h = z$ when the bead and the plate are in contact. 

The overall trajectory of the bead consists of a series of collisions with the plate, at time $t_{n}$ (where $n$ is the index of the collision), separated by a time $\Delta t_{n} \equiv t_{n+1}-t_{n}$ of free fall.
The dissipative nature of the collisions is accounted for by introducing a single parameter, the {\it restitution coefficient} $e$, such that the rebound velocity of the bead (its velocity relative to the velocity of the plate after the collision) is proportional to the impact velocity (the bead velocity relative to the velocity of the plate before the collision), i.e.
\begin{equation}
u_{n}^{+} - v_{n} = - e\,( u_{n}^{-} - v_{n} ).
\label{eq:collision}
\end{equation}
In Eq.~(\ref{eq:collision}), $v_{n}$ stands for the velocity of the plate at the collision $n$, and $u_{n}^{-}$ and $u_{n}^{+}$ for the velocity of the bead before and after the collision, respectively. Note that we thus assume that the collisions are intantaneous and
that $e$ is independent from the impact velocity $u_{n}^{-} - v_{n}$, which is a good approximation for most of the practical cases \cite{Tillett54,Falcon98,Chastaing15}. 

Between two subsequent collisions, during the free fall from time $t_{n}$ to time $t_{n+1}$, the bead is only subject to the acceleration due to gravity $g$ and its height given by
\begin{equation}
h(t) = h_{n} + u_{n}^{+} (t - t_{n}) - \frac{g}{2}\,(t - t_{n})^2
\label{eq:free fall}
\end{equation}
where $h_{n} \equiv h(t_{n}) = z(t_{n})$ is the height at $t = t_n$.

Provided the height of the plate $z(t)$ is known at all times, one can determine the entire trajectory of the bead by using the equations accounting for the collisions [Eq.~(\ref{eq:collision})] and for the free fall [Eq.~(\ref{eq:free fall})]. Indeed, from the knowledge of the takeoff velocity of the bead, $u_{n}^{+}$ right after the collision $n$ at time $t_n$, one can determine the next collision time $t_{n+1}$ from Eq.~(\ref{eq:free fall}) and the condition $h(t_{n+1}) = z(t_{n+1})$. The velocity of the bead before the collision at $t_{n+1}$ is then given by
$u_{n+1}^{-} = u_{n}^{+} - g\,(t_{n+1} - t_{n})$.
Then, one can use the collision rule Eq.~(\ref{eq:collision}) to calculate the takeoff velocity of the bead, $u_{n+1}^{+}$ right after the collision $n+1$. The procedure can be iterated to determine the entire trajectory \cite{Holmes82}.


The aim of our study is to report on statistical properties of the ball trajectory in a chaotic regime. A common way to achieve a chaotic regime consists in imposing a sinusoidal plate-motion, $z(t) = A \sin{(\omega t)}$, with a large dimensionless acceleration $\Gamma=A \omega^2/g$ \cite{Holmes82,Everson86}. However, even if a large $\Gamma$ insures that the bead experiences a chaotic motion, unwanted synchronization effects still remain \cite{Chastaing15}. Techniques to avoid these effects could consist in using quasi-periodic \cite{Oliveira97} or random motion of the plate \cite{Burkhardt06,Garcia-cid15}. Quasi-periodic motion of the plate does not make the underlying frequency disappear and there is still a risk of partial synchronization. The random motion of plate is achieved, in practice, by feeding an electromagnetic, mechanical, system with a random driving signal (a colored Gaussian noise in \cite{Garcia-cid15}) which is then
filtered by the mechanical system. Even if the motion of the plate is characterized subsequently, it is difficult to impose all the characteristics of the plate motion. We proposed lately an alternative protocol \cite{Chastaing15} that gets rid of some of the ball-plate synchronization effects, while keeping constant the maximum velocity $A\,\omega$ reached by the plate during each of the cycles.
The motion of the plate consists of a series of sinusoidal cycles, 
$z_i(t) = A_i \sin{(\omega_i t)}$ such that $A_i\,\omega_i = A\,\omega~(\forall i)$. The cycle $i$ begins at time $t_i$ such that $z_i(t_i)=0$ and lasts for a duration $2\pi/\omega_i$.
This choice insures that the vertical position of the plate $z$, its velocity, $v$, and its acceleration are continuous at all time, whatever the choice of the set of $\omega_i$. The plate motion is thus well-characterized ---not only in the model, but also in its experimental counterpart.

From now on, in order to avoid temporal synchronization of the bead motion with the motion of the plate, the $\omega_i$ are chosen randomly, with uniform probability, in an interval of width 25\% around the central value $\omega$, namely $\omega_i \in [\frac{7}{8}\omega,\frac{9}{8}\omega]$.
We remind that the series of amplitude $A_i$ is accordingly chosen so as to insure $A_i\,\omega_i = A\,\omega~(\forall i)$. In addition, even if the maximum acceleration $A_i\,\omega_i$ is not constant from one cycle to another, we use, for convenience, the dimensionless acceleration
$\Gamma \equiv A\,\omega^2/g$ as the control parameter specifying the working conditions.

\section{Statistical characterization of the trajectory}
\label{sec:statchar}

\begin{figure}[h]
\begin{center}
\includegraphics[width=.45\columnwidth]{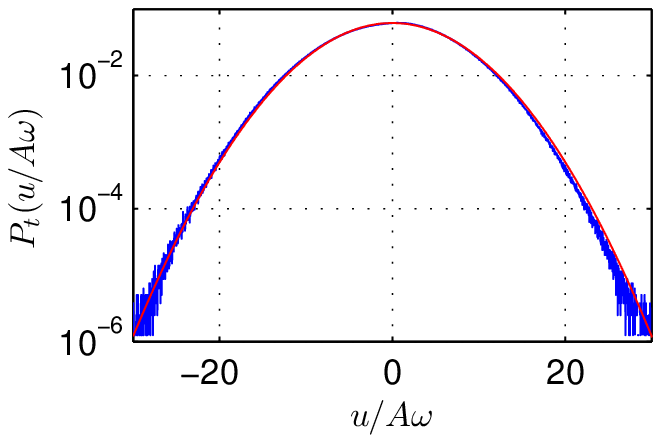}
\includegraphics[width=.45\columnwidth]{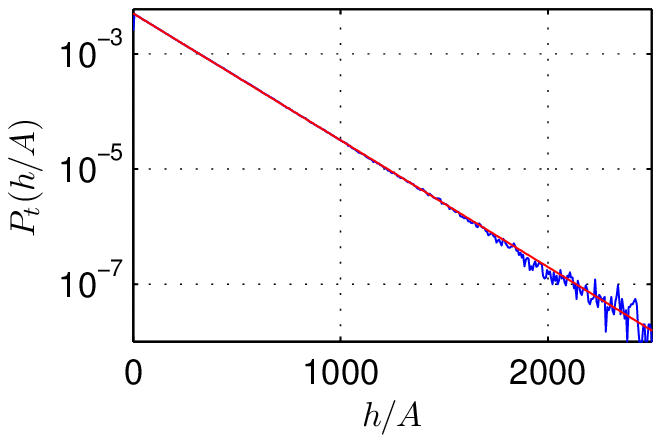}
\end{center}
\caption{\label{fig:pdf_h_u} Distributions $P_t(u)$ of the bead velocity and $P_t(h)$  of the bead height $h$ obtained by averaging over time. Numerical data are shown in blue, while the red lines are interpolations with Eqs.~(\ref{eq:Ptu}) and (\ref{eq:Pth}), respectively ($e=0.975$, $\Gamma=5$).}
\end{figure}

We first determine the probability distribution function of the velocity $u$ and of the height $h$ of the bead through time. 
We thus consider the probabilities $P_t(u)$ and $P_t(h)$ that, all along the trajectory, the bead moves with the velocity $u$ or be at the height $h$, respectively.
Results are shown in Fig.~\ref{fig:pdf_h_u}.
Note that we use dimensionless quantities, $h/A$ and $u/(A\omega)$. Introducing further the mass of the bead, $m$ (even if it does not play any role in the dynamics), to make the next expressions look more familiar,
 we observe that $P_t(u)$ and $P_t(h)$ are well-described by effective equilibrium distributions \cite{Huang}
\begin{align}
\label{eq:Ptu}
P_t(u) &\propto \exp \left(-\frac{m\,u^2}{2kT} \right)\\
\label{eq:Pth}
P_t(h) &\propto \exp \left(-\frac{m\,g\,h}{kT} \right)
\end{align}
with the same temperature $kT$ for both height, $h$, and velocity, $u$. The result holds true as long as the restitution coefficient $e$ remains close to unity.

\begin{figure}[h]
\begin{center}
\includegraphics[width=.45\columnwidth]{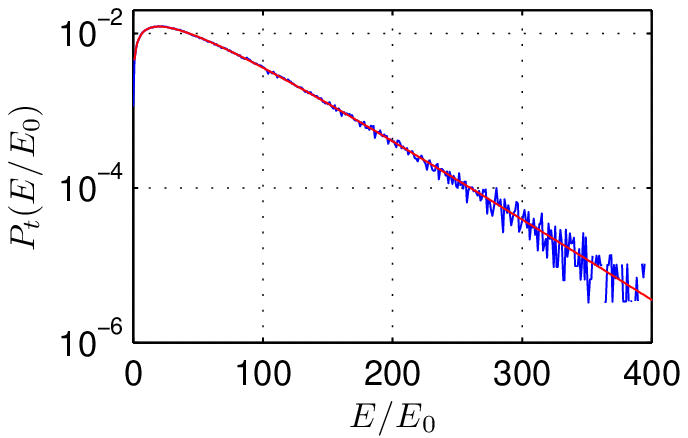}
\includegraphics[width=.43\columnwidth]{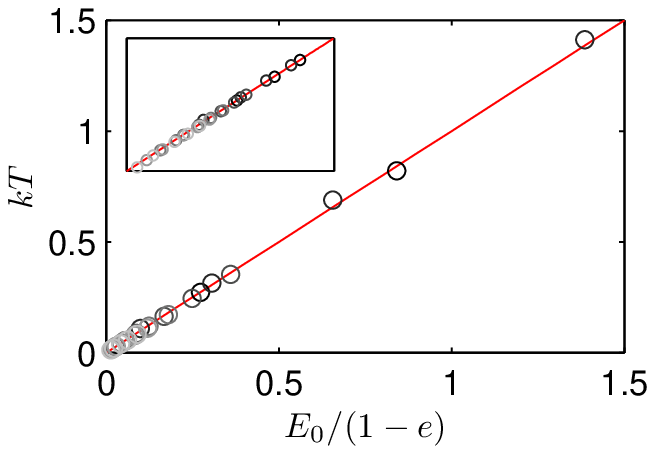}
\end{center}
\caption{\label{fig:pdf_E} Left: Distribution $P_t(E)$ of the total energy $E$ through time. Numerical data are shown in blue, while the red line is the interpolation with Eq.~(\ref{eq:PtE}) ($e=0.975$, $\Gamma=5$). Right: Temperature $kT$ vs.~$E_0/(1-e)$ for various values of $(e,\Gamma)$. The darker the symbol is, the closer to unity is the coefficient of restitution. The red line points out equality
($e \in[0.80;0.99]$ and $\Gamma \in [1.34;5]$).}
\end{figure}

From the same set of data, one can compute the distribution of the total energy $E=\frac{1}{2}\,m\,u^2+m\,g\,h$. Results are shown in the left panel of Fig.~\ref{fig:pdf_E}. In accordance with the distributions of $u$ and $h$ [Eqs.~(\ref{eq:Ptu}) and (\ref{eq:Pth})], we expect
\begin{equation} \label{eq:PtE}
P_t(E) \propto \sqrt{E}\exp\left(-\frac{E}{kT}\right).
\end{equation}
We again observe an excellent agreement with the numerical data.
It is then convenient to use the energy distribution $P_t(E)$ to determine the temperature $kT$ for various values of the restitution coefficient $e$ and dimensionless acceleration $\Gamma$. The results are reported in the right panel of Fig.~\ref{fig:pdf_E}.
We observe that the temperature $kT$ is proportional to the typical energy $E_0 \equiv m\,(A\,\omega)^2$, the dependence on $e$ leading to
\begin{equation}
kT= \frac{m (A \omega)^2}{1-e}.
\label{eq:Kt}
\end{equation}
Note that the above result holds true only if the restitution coefficient $e$ is close to $1$,
limit to which we restrict our study.
This is the reason why Eq.~(\ref{eq:Kt}) differs slightly from the prediction of Warr {\it et. al.} in which the restitution coefficient can depart significantly from unity \cite{Warr96}.

\begin{figure}[h]
\begin{center}
\includegraphics[width=.45\columnwidth]{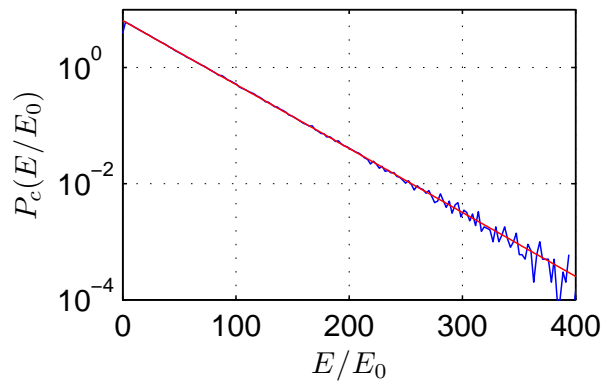}
\end{center}
\caption{\label{fig:pdf_Ec} Distribution $P_c(E)$ through collisions vs. $E/E_0$. Numerical data are shown in blue, while the red line is the interpolation with Eq.~(\ref{eq:PcE}) ($e=0.975$, $\Gamma=5$).}
\end{figure}

We considered, up to now, probabilities of quantities measured all along the trajectory of the bead, through time. 
Alternatively, one may consider probability distributions of quantities at the collisions, instead.
In particular, we report in Fig.~\ref{fig:pdf_Ec} the probability distribution
of the total energy at the collisions, $P_c(E)$, which will prove useful in the next section. 
We observe a clear exponential behavior well accounted for by 
\begin{equation} \label{eq:PcE}
P_c(E) = \frac{1}{kT} \exp\left(-\frac{E}{kT}\right)
\end{equation}
with the same temperature $kT$. The difference between Eq.~(\ref{eq:PtE}) and Eq.~(\ref{eq:PcE}) comes from the reweighting, in the temporal average, of energy value $E$ by the duration of the free fall between two successive collisions. Indeed, the duration of flight associated with larger energy is larger. More precisely, in our working conditions, the maximum height reached by the bead is large compared to the typical amplitude $A$ of the plate motion. In this limit, the duration of the flight is proportional to the takeoff velocity, $u^{+}$. In the same limit, the potential energy is negligible as compared to the kinetic energy and $u^{+} \propto \sqrt{E}$. Taking thus into account that the free-fall duration between two successive collisions is proportional to $\sqrt{E}$, one easily deduces Eq.~(\ref{eq:PcE}) from Eq.~(\ref{eq:PtE}).

In summary, at first sight, the system exhibits probability distributions that greatly resemble equilibrium distributions in spite of the peculiar, and dissipative, nature of the "thermostat". From the probability distributions, one obtains a single temperature $kT$. However, the latter must be taken with care. Indeed, for instance, the temperature that we define here is not a property of the thermostat alone as it depends on the mass, $m$, of the bead ---see Eq.~(\ref{eq:Kt}).
In addition, we have shown that an equilibrium distribution for the energy is also obtained when considering the state of the system at the collisions only, as reported in Fig.~\ref{fig:pdf_Ec}. Doing so, we disregard the fact that the duration of the states depends on the associated energy, which removes the factor $\sqrt{E}$ in front of the exponential in the energy distribution. However, from a formal point of view, we can regard the statistics of energy states separated by contacts of the system (the bead) with the thermostat. 
In the following, we shall consider statistical properties of quantities at the collisions, in particular the energy increments, to discuss the quality of the "thermostat" and the potential definitions of the temperature.

\section{Energy exchanges}
\label{sec:energy exchanges}

We have shown, by studying the distributions of the velocity $u$, height $h$ and total energy $E$, that the bouncing bead could be described, to a good approximation, as an equibrium system at an effective temperature $kT$ (Sec.~\ref{sec:statchar}).
A natural question is thus to know whether the vibrating plate, which both injects and dissipates energy, indeed plays the role of a thermostat at temperature $kT$.
In this section, we first characterize the statistics of energy increments, and then proceed to a more detailed analysis in terms of probability fluxes, with the aim to test the hypothesis of detailed balance that, in principle, underlies the notion of energy reservoir.

\subsection{Statistics of energy exchanges}
\label{sec: stat energy exchanges}

At each collision, the total energy of the bead evolves from a value $E$, before the collision, to $E'$, after the collision. 
The collision occuring at constant height, the variation is only due to the change in the kinetic energy, such that $E'-E \equiv \Delta E=\frac{m}{2}\bigl[(u^+)^2 - (u^-)^2 \bigr]$.

\begin{figure}[h]
\begin{center}
\includegraphics[width=0.45\columnwidth]{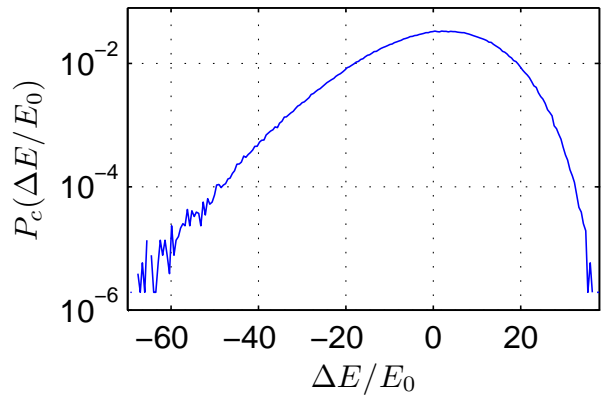}
\end{center}
\caption{\label{fig:pdEt} Probability distribution $P_c(\Delta E)$ of the increments of energy $\Delta E$, obtained by averaging over collisions ($e=0.975$, $\Gamma=5$).}
\end{figure}

Energy exchanges can be characterized by reporting the probability distribution $P_c$
of the increments $\Delta E$ of the total energy $E$ through the collisions.
Results are reported in Fig.~\ref{fig:pdEt}. We first notice that the distribution is asymmetric, which is already a clue that the system is not in contact with a classical thermostat. The most probable value is positive but, we remind, the mean value is zero since, on average, the bead neither loses nor gains energy. The left tail of the distribution is essentially exponential, while its right tail decays faster than exponentially. Such a distribution of energy increments is reminiscent of the BHP distribution reported, among others, in closed turbulent flows \cite{Bramwell98,Bramwell01}.

\subsection{Probability flux between configurations}
\label{sec:flux}

To analyze more precisely the dynamics of energy exchanges, we turn to a detailed characterization of the transition probabilities between different energies.
Our goal is to test whether the energy-exchange dynamics obeys microreversibility (or detailed balance in the language of stochastic processes), a standard requirement for a microscopic heat reservoir \cite{Sokolov}. There is {\it a priori} no reason that such a macroscopic and dissipative system like a bead on a vibrating plate satisfies this property. However, we wish to assess how far a simple macroscopic system can be analyzed and modeled in terms of concepts coming from statistical physics of microscopic systems and thermodynamics.

We first define $\mathcal{P}(E \rightarrow E')$, the probability for the system to evolve from energy $E$ to $E'$ during a collision with the plate. This quantity can be directly measured in the numerical simulations, and will play a key role in the following analysis. To be more specific, $\mathcal{P}(E \rightarrow E')$ is a probability density, in the sense that $\mathcal{P}(E \rightarrow E')dE dE'$ is the probability to observe a transition from the energy interval $[E,E+dE]$ to the energy interval $[E',E'+dE']$.
Detailed balance (or microreversibility) corresponds to the fact that the probability to observe a given trajectory is equal to the probability to observe the time-reversed trajectory. In other words, $\mathcal{P}(E \rightarrow E')=\mathcal{P}(E' \rightarrow E)$.
This property can be explicitly tested in the numerical simulations.

\begin{figure}[h]
\begin{center}
\includegraphics[width=0.40\columnwidth]{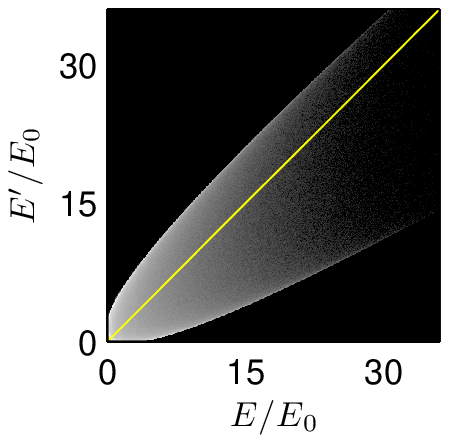}
\hspace{0.5cm}
\includegraphics[width=0.42\columnwidth]{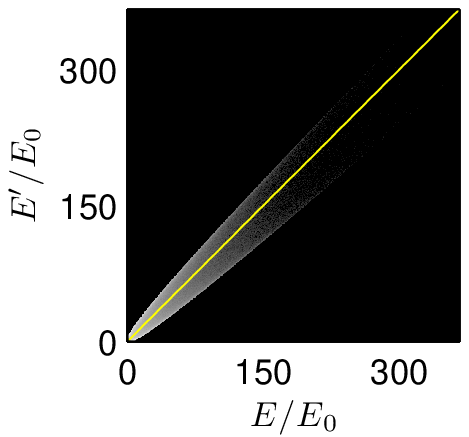}
\end{center}
\caption{\label{fig:map} Maps of the probability $\mathcal{P}$ of the energy of the system to evolve from $E$ to $E'$ in a single collision with the plate.
Values are converted to gray levels (logarithmic scale), the larger values being brighter. The yellow line is the diagonal (Left: $e=0.85$, $\Gamma=4$; Right: $e=0.975$, $\Gamma=5$).}
\end{figure}

We report, in Fig.~\ref{fig:map}, maps of the probability $\mathcal{P}(E \rightarrow E')$ in the $(E,E')$-plane, the value of the probability $\mathcal{P}(E \rightarrow E')$ being converted into gray levels.
%
%
%
%
Detailed balance would manifest itself on these maps as a symmetry about the diagonal. The absence of such symmetry indicates clearly that detailed balance is not satisfied.

\begin{figure}[t]
\begin{center}
\includegraphics[width=0.40\columnwidth]{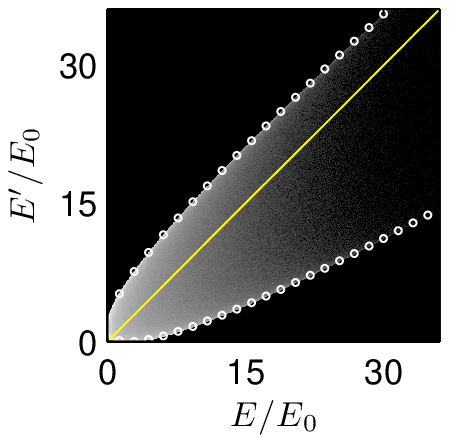}
\hspace{0.5cm}
\includegraphics[width=0.40\columnwidth]{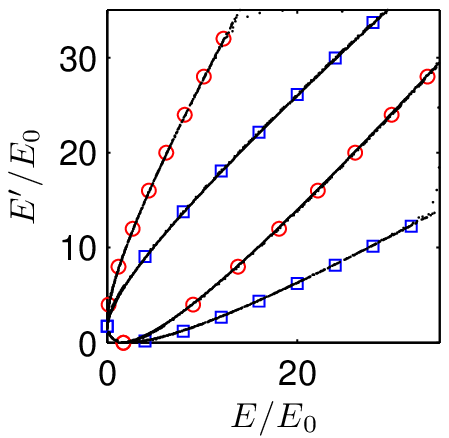}
\end{center}
\caption{\label{fig:mapbis} (Left) Map of the probability $\mathcal{P}$ of the energy of the system to evolve from $E$ to $E'$ in a single collision. Values are converted to gray levels (logarithmic scale), the larger values being brighter.
White circles are from Eqs.~(\ref{eq:Emin}) and (\ref{eq:Emax}).
The yellow line is the diagonal.
(Right) Maxima of $\mathcal{P}$ (blue squares), theoretical curves $E'_{\rm min}(E)$ and $E'_{\rm max}(E)$ (continuous lines) and their symmetrics (red circles and continuous lines) about the diagonal (Left: $e=0.85$, $\Gamma=4$).
}
\end{figure}

Besides, one clearly sees that the transitions occur in a delimited region of the energy map.
Interestingly, this region is delimited by two curves $E'_{\rm max}$ and $E'_{\rm min}$ that can easily be determined analytically, considering that the maximal and minimal velocities of the bead after collisions are obtained from plate velocities equal to $v=A\omega$ and $v=-A\omega$ respectively.
A simple calculation then leads to
\begin{eqnarray}
\label{eq:Emax}
E'_{\rm max} &=& e^2 E + e(1+e)\sqrt{2E E_0}+\frac{1}{2}(1+e)^2 E_0 \\
\label{eq:Emin}
E'_{\rm min} &=& e^2 E - e(1+e)\sqrt{2E E_0}+\frac{1}{2}(1+e)^2 E_0
\end{eqnarray}
When Eqs.~(\ref{eq:Emax}) and~(\ref{eq:Emin}) are reported on the probability map (Fig.~\ref{fig:mapbis}), one clearly sees that all the transitions indeed occur between these two limiting curves. Moreover, we observe that, for a given value of $E$, the two local maxima of the probablity $\mathcal{P}$ correspond to $E'_{\rm min}(E)$ and $E'_{\rm max}(E)$. Indeed, transitions from $E$ to $E'_{\rm min}(E)$ and $E'_{\rm max}(E)$ are more frequent because they correspond to collisions occuring for the minimal or maximal velocity of the plate. Due to the sinusoidal vertical motion, the plate spends more time in these regions of extremal velocity which are associated with the minimum acceleration.

\subsection{Injected and dissipated energy increments}
\label{sec:inject-dissip}

The presence of irreversible transitions suggests that, not surprisingly, the dissipation of energy due to the inelastic collisions plays an important role.
A natural idea is then to decompose the energy exchange $\Delta E=E'-E$ during a collision into dissipated, $E_{\rm diss}$, and injected, $E_{\rm inj}$, energy increments, satisfying
$\Delta E = E_{\rm inj} - E_{\rm diss}$.
Although physically motivated, this decomposition is actually not unique, as we explain below. It has to obey two basic requirements: (i) the dissipated energy increments are always positive, $E_{\rm diss} \ge 0$; (ii) the dissipated energy increments reduce to  $E_{\rm diss} = (1-e^2) E$ in the limit of zero driving.
A popular form of the dissipated energy increments is $E_{\rm diss} = \frac{1}{2} (1-e^2) (u^{-}-v)^2$, that is the dissipated energy in the moving frame of the plate.
However, we are looking for the dissipated energy in the frame of the laboratory, and it is not obvious that the dissipated energy should take the same expression in all frames. Other forms of the dissipated energy can indeed be considered.

To provide a guide to the definition of the dissipated energy increments, we actually need to be more specific about the properties we expect from the increments of the injected energy, $E_{\rm inj}$. Our idea here is to try to model energy injection as an equilibrium reservoir. 
A minimal requirement would be that all transitions are reversible. Assuming that such transitions (associated to injection only) can be represented by a map qualitatively similar to the ones shown in Fig.~\ref{fig:map}, a necessary condition of reversibility is the symmetry of the limiting curves about the diagonal.

We first note that Eqs.~(\ref{eq:Emax}) and (\ref{eq:Emin}) have a nontrivial symmetry that can be revealed as follows. Rewriting these equations into (setting $E_0=1$ for convenience),
\begin{equation} \label{eq:Eminmax}
E' = e^2 E \pm e(1+e)\sqrt{2E}+\frac{1}{2}(1+e)^2,
\end{equation}
where $E'$ stands for $E'_{\rm max/min}$, we can make the following transformation,
\begin{equation}
X=\frac{E'+e^2E}{\sqrt{2}}, \quad Y=\frac{E'-e^2E}{\sqrt{2}}.
\end{equation}
This transformation can be interpreted geometrically as a contraction of the $E$ axis, defining $x=e^2E$ and $y=E'$, followed by a rotation of the $(x,y)$-axes by $45^{\circ}$.
Eq.~(\ref{eq:Eminmax}) then reads, in the variables $(X,Y)$,
\begin{equation} \label{eq:parabola}
X = \frac{\sqrt{2}}{(1+e^2)} \, Y^2 +  \frac{\sqrt{2}}{8} \, (1+e^2) 
\end{equation}
exhibiting a symmetry about the $X$ axis (the curve is invariant by reversing the sign of $Y$).

\begin{figure}[h]
\begin{center}
\includegraphics[width=0.40\columnwidth]{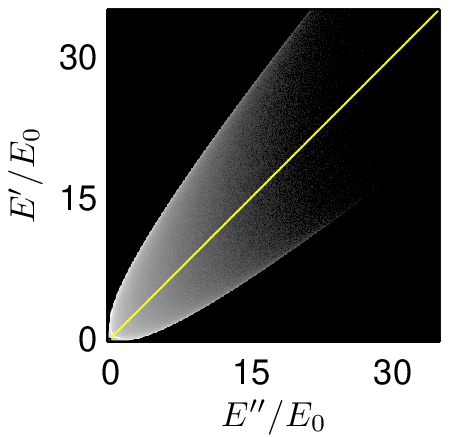}
\hspace{0.5cm}
\includegraphics[width=0.41\columnwidth]{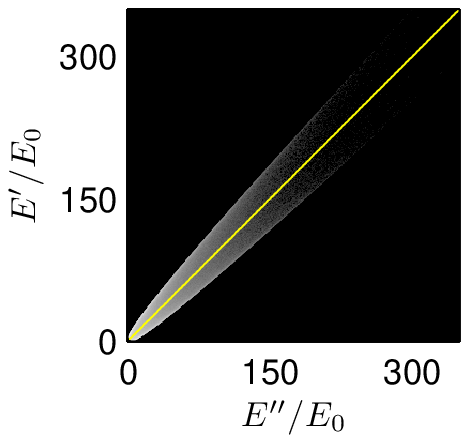}
\end{center}
\caption{\label{fig:map_inj_dis} Map of $\mathcal{P}(E''\equiv e^2E \rightarrow E')$.
Values are converted to gray levels (logarithmic scale), the larger values being brighter.
The yellow line corresponds to the diagonal, as a guide to the eye (Left: $e=0.85$, $\Gamma=4$; Right: $e=0.975$, $\Gamma=5$).}
\end{figure}

This geometrical property suggests the following physical interpretation.
The transition $E \rightarrow E'$ can be decomposed into a dissipation step $E \rightarrow E''$, followed by an injection step $E'' \rightarrow E'$. Such interpretation leads us to define the dissipated energy increments as $E_{\rm diss} = (1-e^2)\,E$, and, consequently, the injected energy as $E_{\rm inj} = E' - e^2\,E$.
This decomposition has the advantage that the probability map of the injected energy increments $E'' \rightarrow E'$ has a symmetric envelope with respect to the diagonal, as a result of the symmetry seen in Eq.~(\ref{eq:parabola}).
This symmetry is confirmed in the maps of $\mathcal{P}(E''\equiv e^2E \rightarrow E')$
reported in Fig.~\ref{fig:map_inj_dis}. The transitions are reversible in the sense that, if a transition from $E''$ to $E'$ is observed, the reverse transition from $E'$ to $E''$ can also be observed. However, the map is not symmetric as a whole, indicating again a violation of detailed balance.

\subsection{Deviation from thermal equilibrium}
\label{sec:temperature}

In this section, we shall use a series of tools developed to characterize the steady-state of out-of-equilibrium systems. We shall see that the partition of the energy exchanges into dissipated and injected increments (Sec.~\ref{sec:inject-dissip}) makes such analysis possible.
We shall also see that each tool leads to a definition of a temperature, whose value will be discussed in Sec.~\ref{sec:discussion}.

\subsubsection{Rate of entropy production}
\label{sec:rate of entropy creation}

First, one can quantify the distance of our system to detailed balance by assessing the rate of entropy production, as defined by Gaspard \cite{Gaspard04} ---see also \cite{Ciliberto07} for experimental measurements of this quantity. 
We focus here on the entropy production associated to the energy injection mechanism, that reads
\begin{equation}
\sigma=\frac{\delta E^2}{2}\sum_{E',E''}~\left[ \mathcal{P}(E'' \rightarrow E')-\mathcal{P}(E' \rightarrow E'') \right]~\ln \left[ \frac{\mathcal{P}(E'' \rightarrow E')}{\mathcal{P}(E' \rightarrow E'')}\right].
\label{eq:entropie_gaspard}
\end{equation}
where $\delta E$ is the width of the bins used to discretize the energy map.
This entropy production can also be considered as a measure of the departure from detailed balance in the injected energy map shown in Fig.~\ref{fig:map_inj_dis}.
Note that $\sigma$ would be ill-defined if one of the transitions, allowed in one direction, was not permitted in the reverse direction. 
This is why an appropriate partition in dissipated and injected increments is necessary. This is also the reason why an entropy production cannot be measured in this framework for the energy dissipation.

\begin{figure}[h]
\begin{center}
\includegraphics[width=0.45\columnwidth]{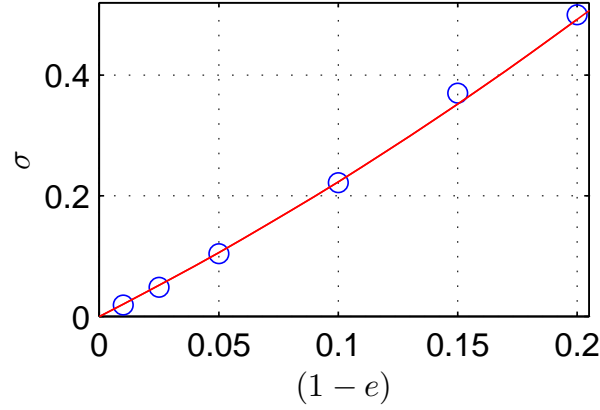}
\end{center}
\caption{\label{fig:sigma} Rate of entropy creation $\sigma$ vs.~$(1-e)$ for several values of ($e$,$\Gamma$). Red line: $\sigma=2\,(1-e) $.
The rate $\sigma$ depends on the restitution coefficient $e$, but not on $\Gamma$. Each point corresponds to several values
of $\Gamma$ (i.e., 1.34, 1.70, 2.35, 2.86, 3.44 , 4.00 and 5.00).
}
\end{figure}

In practice, $\sigma$ depends on the binning in energy $\delta E$. In order to get a value of $\sigma$ corresponding to the continuous limit, we extrapolate the observed linear decrease of $\sigma(\delta E)$ to $\delta E=0$.
We observe in Fig.~\ref{fig:sigma} that $\sigma$, which does not depend on the dimensionless acceleration $\Gamma$, is a function of $(1-e)$ only, vanishing for $e = 1$, according to
\begin{equation}
\sigma =(2.05 \pm 0.05)(1-e)  + (2.3 \pm 0.1)\,(1-e)^2.
\label{eq:entropcreat}
\end{equation}
Thus, the violation of the reversibility is indeed growing for decreasing $e$, i.e., when dissipation is increased.

\begin{figure}[h]
\begin{center}
\includegraphics[width=0.45\columnwidth]{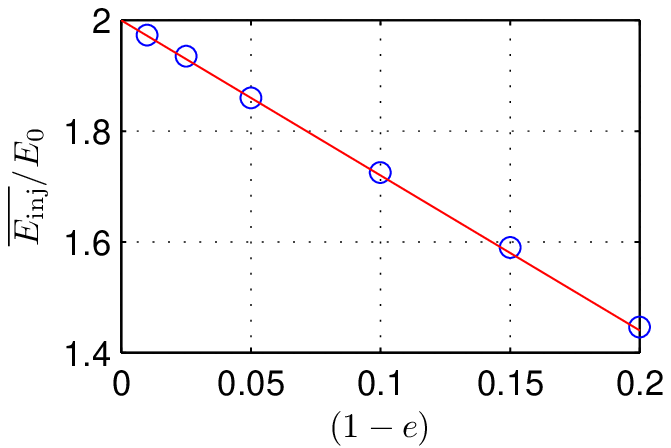}
\includegraphics[width=0.45\columnwidth]{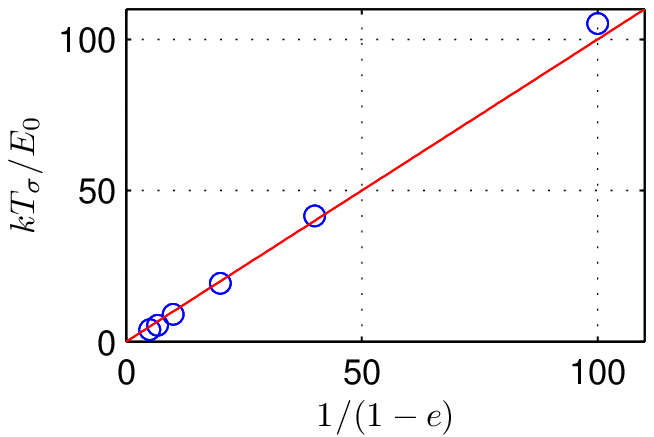}
\end{center}
\caption{\label{fig:kTS} (Left) Mean injected energy $\overline{E_\mathrm{inj}}/E_0$ vs. $(1-e)$. (Right) Temperature $kT_\sigma/E_0$ vs. $1/(1-e)$ for several values of $\Gamma$
(1.34, 1.70, 2.35, 2.86, 3.44 , 4.00 and 5.00). Red line: $kT_\sigma = E_0/(1-e)$.
}
\end{figure}

Interestingly, $\sigma$ can be used to measure a temperature scale. Indeed, from the chosen point of view, the bead exchanges energy in two separated steps, dissipation and injection. 
As already mentioned, we consider here only the injection energy, and not the dissipated one.
Using a heuristic thermodynamic argument, the entropy production satisfies $\sigma/\overline{E_\mathrm{inj}} =1/kT_1-1/kT_2$, where $\overline{E_\mathrm{inj}}$ denotes the average of the injected energy, $kT_1$ being the temperature of the system (playing here the role of the cold thermostat) and $kT_2$ the temperature of the injection mechanism (hot thermostat). Note that, one would need two equations to determine $kT_1$ and $kT_2$ properly. However, we can estimate a typical temperature $kT_{\sigma}=\overline{E_\mathrm{inj}}/\sigma$. From the numerical data (Fig.~\ref{fig:kTS}, left), we get $\overline{E_\mathrm{inj}}=[2.00 -(2.77 \pm 0.05)(1-e)]\,E_0$.
Then, reporting $kT_{\sigma}/E_0$ as a function of $1/(1-e)$ (Fig.~\ref{fig:kTS}, right), we observe that, in the limit $e \to 1$,
\begin{equation} \label{eq:kTsigma}
kT_\sigma = (1.00 \pm 0.02)\,\frac{E_0}{(1-e)},
\end{equation}
very close to the temperature of system $kT$ defined in Sec.~\ref{sec:statchar} from the distribution of the energy.
Among the simple interpretations of this result,
a tentative interpretation would be that the temperature $kT_1$ is equal to the temperature $kT$ measure from the energy distribution, leading to an infinite temperature $kT_2$ for the injection thermostat.
Another simple interpretation could be that the injection mechanism is at a temperature $kT_2=kT$, while the system would be, from the point of view of energy injection, at an effective temperature $kT_1=kT/2$. Of course, infinitely many other values of $kT_1$ and $kT_2$ are also compatible with Eq.~(\ref{eq:kTsigma}). Note also that the relation $\sigma/\overline{E_\mathrm{inj}} =1/kT_1-1/kT_2$ is purely phenomenological here, and should not be taken for granted.

\subsubsection{Transition rates}
\label{sec:Transition rates}

An additional test of detailed balance can be performed by estimating the transition rates $\mathcal{T}$.
Considering the total energy increments, given that the energy distribution $P_c(E)$ is known, one can determine from the knowledge of $\mathcal{P}(E \rightarrow E')$ the transition rate $\mathcal{T}(E \rightarrow E')$ using the relation
\begin{equation} \label{eq:DBbasic}
\mathcal{P}(E \rightarrow E') \equiv P(E) \, \mathcal{T}(E \rightarrow E').
\end{equation}
If detailed balance was holding with respect to the energy distribution $P_c(E)\propto \exp(-E/kT)$, the transition rate $\mathcal{T}$ would satisfy
\begin{equation}
\mathcal{T}(E \rightarrow E')\, \exp(-E/kT) = \mathcal{T}(E' \rightarrow E)\, \exp(-E'/kT)
\end{equation}
which would lead to the relation
\begin{equation} \label{eq:DBcan}
-\ln \left[\frac{\mathcal{T}(E \rightarrow E')}{\mathcal{T}(E' \rightarrow E)}\right] = \frac{1}{kT}\, (E'-E)
\end{equation}
Lifting the assumption of detailed balance, inspired by Eq.~(\ref{eq:DBcan}), one may define the function
\begin{equation}
g(E,E') = -\ln \left[\frac{\mathcal{T}(E \rightarrow E')}{\mathcal{T}(E' \rightarrow E)}\right].
\label{eq:g}
\end{equation}
If the detailed balance was holding, a scatter plot of $g(E,E')$ as a function of the difference $E'-E$ would lead, according to Eq.~(\ref{eq:DBcan}), to a straight line of slope $1/kT$.

\begin{figure}[h]
\begin{center}
\includegraphics[width=0.47\columnwidth]{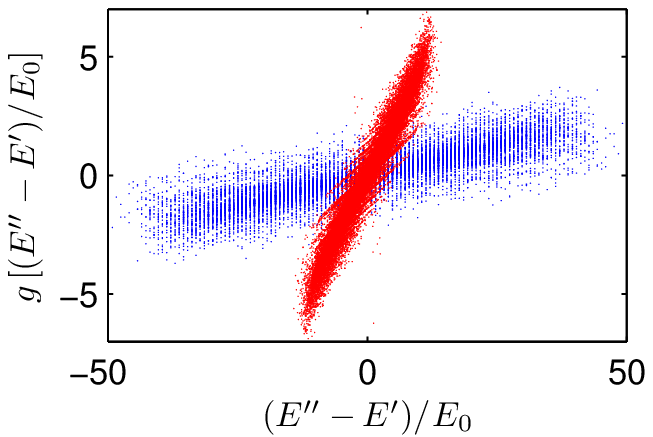}
\includegraphics[width=0.45\columnwidth]{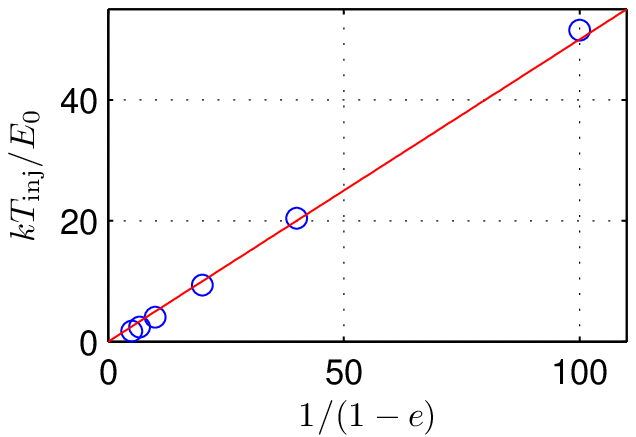}
\end{center}
\caption{\label{fig:kT_inj} Function $g(E'',E')$ vs.~increment $(E''-E')/E_0$. Note that the points reasonably align along straight lines [red dots: $e=0.85$, $\Gamma=4$, blue dots: $e=0.975$, $\Gamma=5$] -- Temperature $kT_{\rm inj}/E_0$ vs. $1/(1-e)$ (blue circles) for several values of ($e$,$\Gamma$). Red line: $kT_{\rm inj} = 0.5\,E_0/(1-e)$
($\Gamma \in [1.34 ; 5]$ and $e \in [0.8;0.99]$).}
\end{figure}

Because of the existence of irreversible transitions (Fig.~\ref{fig:map}, Sec.~\ref{sec:flux}), $g(E,E')$ is ill-defined for the increments of the total energy. However, by construction, the map of the injected increments possesses the adequate symmetry and, by analogy with Eq.~(\ref{eq:g}), one can estimate $g(E'',E')$, where $E''=e^2 E$.
In case the detailed balance was holding, $g(E'',E')$ would obey the relation
\begin{equation}
g(E'',E') = \frac{1}{kT_{\rm inj}} (E'-E''),
\end{equation}
where $kT_{\rm inj}$ is an effective temperature characterizing the injection reservoir. Note that $kT_{\rm inj}$ may differ from $kT$ due to the presence of dissipation in addition to the injection mechanism.
We report in Fig.~\ref{fig:kT_inj} the scatter plot of $g(E''E')$ as a function of $E'-E''$.
We observe that points accumulate close to a mean straight line, but with a finite transverse extension. Hence, strictly speaking, the detailed balance does not hold as anticipated from the map shown in Fig.~\ref{fig:map_inj_dis}, but the presence of a mean line around which points accumulate suggests an interpretation of the results as a 'blurred' detailed balance relation. Although this is not fully justified on theoretical ground, we propose to use the slope of the best linear interpolation as an effective temperature $kT_{\rm inj}$.
Measuring $kT_{\rm inj}$ for different values of the restitution coefficient $e$, we empirically obtain
\begin{equation} \label{eq:kTinj}
kT_{\rm inj} = (0.50 \pm {0.02})\,\frac{E_0}{(1-e)}.
\end{equation}
A surprising result is that the temperature $kT_{\rm inj}$ is found to be smaller than (the half of!) the temperature $kT$ characterizing the distributions of velocity and height of the bead (Sec.~\ref{sec:statchar}).
Although we do not have a clear explanation for this measured value of the injection temperature, we discuss in appendix~\ref{appendix} how the temperature measured by fitting a ``noisy'' detailed balance relation can significantly differ from the ``true'' value of the injection temperature.
Hence the result given in Eq.~(\ref{eq:kTinj}) does not rule out the possibility that the ``true'' injection temperature is higher than the system temperature.

\subsubsection{Fluctuation relations}

To gain further insight on the injection process, we now analyze the fluctuations of injected energy from the point of view of fluctuation relations (Gallavotti-Cohen, or Crooks, relation) \cite{Gallavotti95,Kurchan98,Lepri98,Aumaitre01,Ciliberto04}.
We introduce the mean injected power during $n$ successive impacts
\begin{equation}
\dot{w}_{n}(i) = \frac{\sum_{k=i}^{i+n} E_{\rm inj}(k)}{(t_{i+n}-t_i)}
\label{eq:wdot}
\end{equation}
Note that the theorem is generally expressed using time averaging. Here, in the sake of continuity, we use a fixed number $n+1$ of successive impacts, instead of a fixed time window.
The fluctuation relation gives some characteristics of the probability of events $\dot{w}_{n}(i)$. It assumes, in particular, that for large enough values of $n$, there exists a function $f$ such that $P(\dot{w}_{n})=\exp \left[ -n \tau f(\dot{w}_n) \right]$. 
Then a fluctuation relation holds if the asymmetry function $\delta(\dot{w}_n) \equiv f(-\dot{w}_n)-f(\dot{w}_n)$ obeys a linear relation, namely:
\begin{equation}
\delta(\dot{w}_n) \equiv \frac{1}{n \tau} \ln \Bigl[ \frac{P(\dot{w}_n)}{P(-\dot{w}_n)} \Bigr] = \beta_{\rm GC} \, \dot{w}_n \, , \qquad n \to \infty.
\label{eq:GC2}
\end{equation}
A temperature associated to this fluctuation relation may be defined as
\begin{equation}
kT_{\rm GC} \equiv \frac{1}{\beta_{\rm GC}}.
\label{eq:TGC}
\end{equation}
Note that we cannot strictly reach this limit since we have finite statistics. However, we measured for each set of data associated to the parameters $(e,\Gamma)$, the correlation number of impacts $n_c$, i.e., the typical number of impacts needed for the system to loose memory. On the one hand, we observe that it is necessary to consider at least $n > 5\,n_c$ collisions to get correct averages. On the other hand, we cannot average over very large number $n$ of collisions, because the width of $P(\dot{w}_n)$ is decreasing with $n$. 
Taking into account these two limitations, we decided to average until the asymmetry function is defined in the range $\dot{w}_n \in [0;5 \overline{E_{\rm inj}}]$. We noticed that we could not reach the long-time limit for  values of $e$ too close to unity ($e > 0.95$). As a consequence, we report  results for $e \in [0.8;0.95]$, only.

\begin{figure}[h]
\begin{center}
\includegraphics[width=0.47\columnwidth]{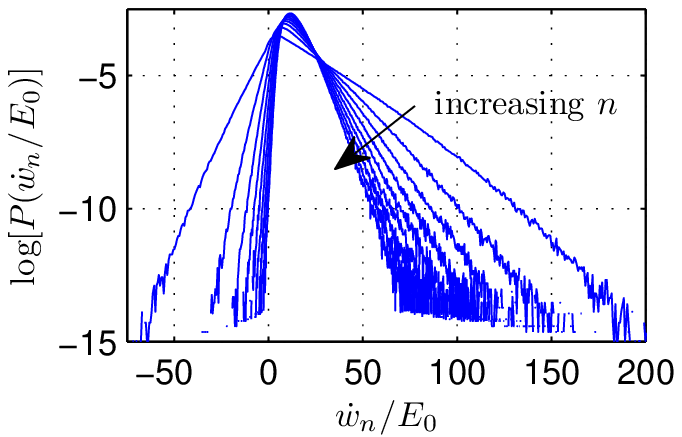}
\includegraphics[width=0.45\columnwidth]{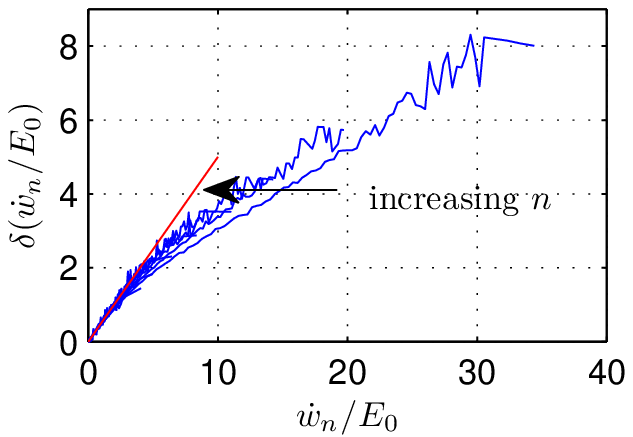}
\end{center}
\caption{\label{fig:GC} (left) $\ln[P(\dot{w}_n/E_0)]$ and (right) asymmetry function $\delta(\dot{w}_n/E_0)$ for several values of $n$ ($e=0.90$, $\Gamma=4$ and $n \in [5;45]$).
Red line: slope at the origin.}
\end{figure}

\begin{figure}[h]
\begin{center}
\includegraphics[width=0.45\columnwidth]{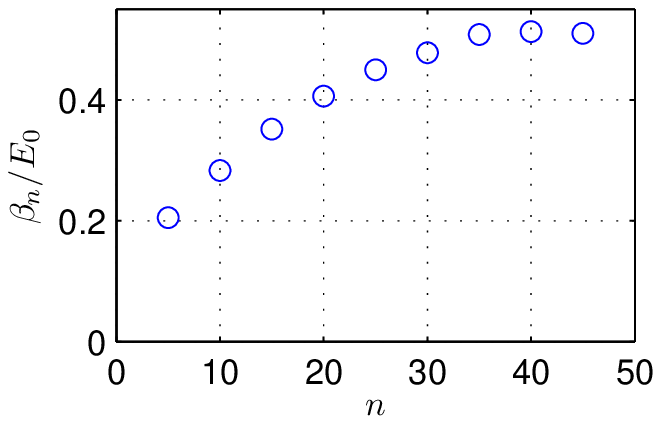}
\includegraphics[width=0.45\columnwidth]{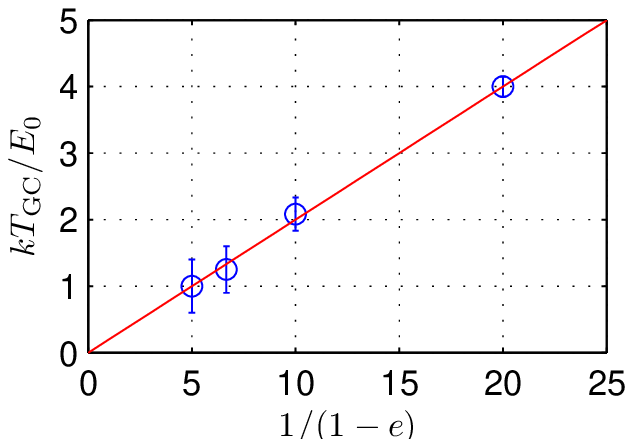}
\end{center}
\caption{\label{fig:GC_2} (Left) $\beta_n/E_0$ as function of $n$ [$e=0.90$, $\Gamma=4$], (Right) Temperature $kT_{\rm GC}/E_0$ vs. $1/(1-e)$ obtained for several values of ($e$,$\Gamma$). Red line : $kT_{\rm GC} = 0.2\,E_0/(1-e)$ ($\Gamma \in [2.35 ; 5]$ and $e \in [0.8;0.90]$). 
}
\end{figure}

In Fig.~\ref{fig:GC}, we report examples of the distribution $P(\dot{w}_n)$ and the associated asymmetry function $\delta(\dot{w}_n)$.
We observe that the fluctuation relation is not satisfied in its standard form Eq.~(\ref{eq:GC2}), as $\delta(\dot{w}_n)$ is not linear in the injected power $\dot{w}_{n}$.
However, we observe, by increasing the number $n$ of successive impacts considered in the average, that the width of the distribution $P$ decreases and that the asymmetry function evolves toward a linear law.
In Fig.~\ref{fig:GC_2}, we report the slope of the asymmetry function at the origin ($\dot w_n \to 0$) as a function of $n$. We observe that the latter slope tends to a constant value for $n \simeq 45$. We then get an estimate of the temperature $kT_{\rm GC}$ which is defined as the inverse of the slope [Eq.~(\ref{eq:TGC})].
Reproducing this protocol for several values of $(e,\Gamma$) and reporting the data as a function of $1/(1-e)$ (Fig.~\ref{fig:GC_2}), we get
\begin{equation}
kT_{\rm GC} =(0.20 \pm {0.02})\,\frac{E_0}{(1-e)}.
\end{equation} 
This temperature is again proportional to $E_0/(1-e)$, alike $kT$, $kT_\sigma$
and $kT_{\rm inj}$. 

\section{Discussion and conclusion}
\label{sec:discussion} 

In this work, we aimed at analyzing a very simple driven mechanical-system, a bead bouncing on a vibrated plate, using concepts from equilibrium and nonequilibrium statistical physics.
We first observed that the distributions of the velocity and height of the bead, as well as of the total energy, follow a simple equilibrium form, with an effective temperature $kT$ 
which depends in a simple way on the physical parameters characterizing, on the one hand, the driving by the vibrating plate and, on the other hand, the energy dissipation during inelastic collisions
between the bead and the plate.
Note that this temperature additionally depends on the mass of the particle, unlike the temperature of standard thermostats.

However, a more careful and detailed analysis reveals that this, apparently, simple effective-equilibrium picture actually hides a complex nonequilibrium dynamics
at the level of the energy exchanges between the bead and the plate.
A simple analysis in terms of a single reservoir at temperature $kT$ fails: indeed,  the reservoir does not follow standard requirements of reversibility of the transition,
in the sense that some transitions between two energy values that are possible in one direction are not possible in the reverse direction.
This feature alone already prevents any possibility of detailed balance.

We show that reversibility can be partly recovered by separating a dissipative contribution from the energy exchanges between the bead and the plate.
Provided an appropriate definition of the increments of dissipated energy, we restore the reversibility of the increments of injected energy in the sense that all the transitions 
that are possible in one direction are possible in the reverse direction. Such procedure makes possible to perform tests on the statistical properties of the energy increments.

First, one can assess the rate of entropy production (Sec.~\ref{sec:rate of entropy creation}) and, then, using the relation with the heat flux across the system, define a first temperature characterizing the system $kT_\sigma$, which is thus found to be of the order of $kT$.
Second we find that strict detailed balance is violated but that, making use of the partial symmetry of the probability map, we reveal that the data show a trend toward a 'blurred' version of detailed balance.
The temperature, $kT_{\rm inj}$ extracted from this approximate detailed balance relation is proportional, but not equal, to the  temperature $kT$ of the bead. More surprizing, while one would expect, due to the presence of dissipation, that an injection reservoir would be characterized by a temperature  higher than that of the bead, $kT_{\rm inj} \simeq 0.5\,kT$ is found to be significantly lower than $kT$.
We propose a tentative way out of this conundrum by showing in appendix \ref{appendix} that the breaking of detailed balance also induces a bias in the temperature measurement that could account for the paradoxical result obtained. Further research in this direction would actually be needed to confirm this potential explanation. 
Finally, we complement the analysis of the data by considering briefly another --now standard-- way of characterizing an energy injection source, namely the fluctuation relations.
Even if the fluctuation relation in its standard form is not satisfied, it is still possible to extract a temperature scale from the data, leading to a temperature value $T_{\rm GC} \simeq 0.2\,kT$.

We see that even one of the simplest driven macroscopic system one can think of can hardly be described in details by equilibrium statistical physics concepts, despite the equilibrium form of the distribution of velocity and height of the bead. Note that, in spite of the proliferation of their values, we find that all temperatures scale in the same way, meaning that the ratios $kT_{\sigma}/kT$, $kT_{\rm inj}/kT$ and $kT_{\rm GC}/kT$ are all independent of the acceleration $\Gamma$ and of the restitution coefficient $e$. However, we observe that the precise value of the temperature depends on the tool we use, which is not a surprize as they do not characterize exactly the same physical processes.  

To conclude, let us emphasize that the fact standard equilibrium concepts do not apply in this strongly driven system was expected and does not come as a surprise. What is actually more surprising is that these concepts at least partially apply, and still provide reasonable guidelines to study the system. Indeed, the probability distributions of velocity, height and energy take an equilibrium form
with an effective temperature, and a ``noisy'' version of detailed balance holds once a suitable splitting between dissipated and injected energy is made. The fact that not all temperatures measured coincide is certainly the price to pay in order to describe such a nonequilibrium systems with close-to-equilibrium concepts. Future work should probably try to identify a theoretical framework which possibly unifies the description using genuine nonequilibrium concepts.

\appendix
\section{Measuring temperature without detailed balance}
\label{appendix}

In this appendix, we discuss how the lack of detailed balance may lead to an effective temperature, measured from a 'blurred' detailed balance relation as shown on Fig.~\ref{fig:kT_inj}, that is different from the 'nominal' temperature at which degrees of freedom are seen to equilibrate.

Let us consider a nonequilibrium reservoir such that a system put in contact with it equilibrates at a temperature $T_0$.
In the absence of detailed balance, a nonzero probability flux, 
\begin{equation}
J(E,E') = \mathcal{P}(E \rightarrow E') - \mathcal{P}(E' \rightarrow E),
\end{equation}
is present between energies $E$ and $E'$.
The probabilities $\mathcal{P}(E \rightarrow E')$ obey the relation
\begin{equation}
\mathcal{P}(E \rightarrow E') = \mathcal{T}(E \rightarrow E')\, C \exp(-E/kT_0)
\end{equation}
where $C\, \exp(-E/kT_0) = P(E)$ is the energy probability distribution.
A straightforward calculation leads to
\begin{equation}
g(E,E') = -\ln \frac{\mathcal{T}(E \rightarrow E')}{\mathcal{T}(E' \rightarrow E)} = \frac{E'-E}{kT_0} - \ln \left[ 1+\frac{J(E,E')\, \exp(E'/kT_0)}{C\, \mathcal{T}(E' \rightarrow E)} \right].
\end{equation}

Let us now assume that the scatter plot of $g(E,E')$ remains reasonably close to a straight line of slope $1/kT_{\rm eff}$.
The effective temperature $T_{\rm eff}$ is basically measured from the scatter plot by computing a (weighted) average of the type

\begin{equation} \label{eq:corrected-T}
\frac{1}{kT_{\rm eff}} = \left< \frac{g(E,E')}{E'-E} \right> = \frac{1}{kT_0} - \left< \frac{1}{E'-E} \ln \left( 1+\frac{J(E,E')\, \exp(E'/kT_0)}{C\, \mathcal{T}(E' \rightarrow E)} \right) \right>
\end{equation}
which immediately shows that the effective temperature $T_{\rm eff}$ measured from the scatter plot of $g(E,E')$ {\it a priori} differs from the temperature $T_0$.
The corrective term has a complicated expression from which it is difficult to guess its sign.
However, the above calculation suggests the following scenario to understand the surprising result found in Sec.~\ref{sec:inject-dissip} that the injection reservoir has an effective temperature $T_{\rm inj}$ that is smaller than the temperature $T$ at which the bead equilibrates.
According to Eq.~(\ref{eq:corrected-T}), it would be possible that the injection reservoir has a nominal temperature $T_0 >T$, but that the effective temperature $T_{\rm inj}$ measured from the scatter plot of $g(E,E')$ is smaller than $T_0$. Although this scenario is speculative, it offers a tentative way to reconcile our observations with the intuition that the temperature of the reservoir should be larger than $T$, given the additional presence of dissipation.

\end{document}